\documentclass[a4paper]{jpconf}
\usepackage{graphicx}
\usepackage{amsfonts}
\usepackage{amsmath,amssymb,mathrsfs}
\begin{document}
\title{A gauged baby Skyrme model and a novel BPS bound}

\author{C Adam$^1$, C Naya$^1$, J Sanchez-Guillen$^1$ and A Wereszczynski$^2$}

\address{$^1$ Departamento de F\'isica de Part\'iculas, Universidad
de Santiago, and Instituto Galego de F\'isica de Altas Enerx\'ias
(IGFAE) E-15782 Santiago de Compostela, Spain}

\address{$^2$Institute of Physics,  Jagiellonian University,
Reymonta 4, Krak\'{o}w, Poland}

\ead{adam@fpaxp1.usc.es, carlos.naya87@gmail.com, joaquin@fpaxp1.usc.es, wereszczynski@th.if.uj.edu.pl}

\begin{abstract}
The baby Skyrme model is a well-known nonlinear field theory supporting topological solitons in two space dimensions. 
Its action functional consists of a potential term, a kinetic term quadratic in derivatives (the "nonlinear sigma model term") and the Skyrme term quartic in first derivatives.
The limiting case of vanishing sigma model term (the so-called BPS baby Skyrme model) is known to support exact soliton solutions saturating a BPS bound which exists for this model. Further, the BPS model has infinitely many symmetries and conservation laws. Recently it was found that the gauged version of the BPS baby Skyrme model with gauge group U(1) and the usual Maxwell term, too, has a BPS bound and BPS solutions saturating this bound. This BPS bound is determined by a superpotential which has to obey a superpotential equation, in close analogy to the situation in supergravity. 
Further, the BPS bound and the corresponding BPS solitons only may exist for potentials such that the superpotential equation has a global solution. We also briefly describe some properties of soliton solutions.

\end{abstract}

\section{Introduction}
The Skyrme model was originally introduced by Skyrme as a novel way to describe nucleons and nuclei \cite{skyrme}, and has ever since served both as a paradigmatic nonlinear field theory supporting topological solitons \cite{BaSu1} and as a successful complementary approach to strong interaction physics \cite{AdNaWi}. Concretely, the primary fields in the Skyrme model are the Goldstone bosons of quantum chromodynamics (pions) taking values in the group manifold SU(2), whereas the nucleons and nuclei emerge as collective excitations of the former, i.e., topological solitons. Due to the complexity of the resulting nonlinear field theory, lower-dimensional theories containing similar topological solitons have been studied, too (e.g., the so-called baby Skyrme model in 2+1 dimensions \cite{old}), both as toy models for the 3+1 dimensional theory and as independent objects of study with their proper applications, e.g., in condensed matter physics and in brane cosmology.  

If the action is required to be Poincare invariant and to allow for a proper Hamiltonian formulation (no higher than second powers in time derivatives) then the possible Lagrangians for the Skyrme and baby Skyrme models are, in fact, quite restricted. For the Skyrme model with target space SU(2) it consists (in ascending order of powers of first derivatives) of a potential term, the "non-linear sigma model" term (quadratic), the Skyrme term (quartic in first derivatives) and a "topological" term (the square of the topological or baryon current, sextic in first derivatives). For the baby Skyrme model with target space $S^2$ (two-sphere) we have a potential term, the quadratic non-linear sigma term, and a quartic term which is, at the same time, topological (the square of the topological current). Topological soliton solutions have been found numerically both for the full models and for various restrictions obtained by suppressing some of the terms described above.  

It has been found recently that the specific restrictions to the potential and topological terms only (the so-called BPS Skyrme \cite{BPS-Sk} and BPS baby Skyrme models \cite{rest-bS}) possess both a BPS bound and exact soliton solutions saturating this bound (the full Skyrme models have BPS bounds, too, but solitons in general do not saturate these bounds). In addition, these BPS submodels exhibit the property of generalized integrability and have, therefore, infinitely many symmetries and conservation laws. 

A last relevant piece of information consists in the fact that both the Skyrme \cite{wit1} and baby Skyrme models \cite{schr1} may be coupled to a U(1) gauge field (i.e., to electromagnetism) in a natural way, and that soliton solutions of these gauged Skyrme \cite{pie-tra} and baby Skyrme models \cite{GPS} have been found numerically. This leads to the natural question whether the electromagnetically gauged versions of the BPS submodels share some of the nontrivial properties (saturation of BPS bounds, infinitely many symmetries and conservation laws) with the ungauged theories. It is the purpose of these notes to briefly demonstrate that for the BPS baby Skyrme model this is indeed the case. Specifically, we want to describe a rather novel, nontrivial BPS bound (for more details and a more exhaustive list of references we refer to \cite{gaugedBPSbaby}).   

\section{The gauged BPS baby Skyrme model}

\subsection{Baby Skyrme models}
The fields of the baby Skyrme model form a three-component vector $\vec \phi =(\phi_1 ,\phi_2 ,\phi_3)$ taking values in the unit two-sphere $S^2$, $\vec \phi^2 =1$. The corresponding baby Skyrme Lagrangian is
\begin{equation}
L=E_0 \int d^2 x \left( -\mu^2 V( \vec\phi ) + \frac{\nu^2}{2} (\partial_\alpha \vec \phi )^2 - \frac{\lambda^2}{8} K_\alpha^2   \right) 
\end{equation}
where $E_0$ is a constant with the dimension of energy, and $(\mu ,\nu ,\lambda )$ are constants with the dimensions of $(l^{-1},l^0 ,l^1)$, respectively ($l$ stands for "length"). Here, the first term on the r.h.s. is the potential term, the second one the sigma model term, and the third term is the square of the topological current $
K^\alpha = \epsilon^{\alpha\beta\gamma} \vec \phi \cdot (\partial_\beta \phi \times \partial_\gamma \vec\phi ) .
$
Finite energy field configurations are required to approach a unique value $\vec \phi_0$ at spatial infinity (where necessarily the potential must obey $V(\vec \phi_0)=0$), therefore the base space ${\mathbb R}^2$ is effectively compactified to a $S^2$, and finite energy field configurations may be classified by an integer winding number $k=(1/8\pi) \int d^2 x K^0$. For the potential we now assume that it only depends on $\phi_3$ and that its unique vacuum is at $\vec \phi_0 = (0,0,1)$, i.e., $V=V(\phi_3)$, $V(\phi_3 =1)=0$. This implies that the potential and, therefore, the full Lagrangian, is still invariant under rotations of the field vector $\vec \phi$ about the three axis (i.e. rotations in the $\phi_1$-$\phi_2$ plane). 
A natural coupling to electromagnetism consists in gauging this residual U(1) symmetry. To achieve this, partial derivatives have to be replaced by  covariant derivatives 
$
D_\alpha \vec \phi \equiv \partial_\alpha \vec \phi + A_\alpha \vec \phi_0 \times \vec \phi 
$
(where $A_\alpha$ is the gauge potential, and $F_{\alpha\beta}\equiv \partial_\alpha A_\beta - \partial_\beta A_\alpha$),
and the standard Maxwell term has to be added. The resulting Lagrangian is
\begin{equation}
L=E_0 \int d^2 x \left( -\mu^2 V( \phi_3 )  + \frac{\nu^2}{2} (D_\alpha \vec \phi )^2 - \frac{\lambda^2}{8}\tilde K_\alpha^2  - \frac{1}{4g^2} F_{\alpha\beta}^2 \right) 
\end{equation}
where
$
\tilde K^\alpha = \epsilon^{\alpha\beta\gamma} \vec \phi \cdot (D_\beta \phi \times D_\gamma \vec\phi ) = K^\alpha + 2\epsilon^{\alpha\beta\gamma} A_\beta \partial_\gamma (\vec \phi_0 \cdot \vec \phi) .
$

\subsection{BPS bounds}
From now on we restrict to the BPS submodels, i.e., we set $\nu =0$ in the above Lagrangians. The energy functional of the ungauged BPS submodel is ($q\equiv (1/2)K_0$) \\
$
E = \frac{1}{2} E_0 \int d^2 x \left[ \lambda^2 q^2 +2 \mu^2 V(\phi_3)  \right] ,
$
and obeys the BPS bound
\begin{equation}
E=\frac{1}{2}E_0 \int d^2 x \left( \lambda q \pm \mu \sqrt{2V}\right)^2 \mp E_0 \lambda \mu \int d^2 x q \sqrt{2V} \ge 
\mp E_0 \lambda \mu \int d^2 x q \sqrt{2V}
\end{equation}
with equality if the BPS equation
$
\lambda q \pm \mu \sqrt{2V} =0
$
is satisfied. The bound is genuinely topological (i.e., for a given potential only depends on the winding number $k$), because the two-form $d^2 x q$ is the pullback (under $\vec \phi$) of the target space area two-form $d\Omega$, and, therefore, $d^2 x q \sqrt{2V(\phi_3)}$, too, is the pullback of a two-form on target space \cite{speight1}, whose integral results in
\begin{equation}
\int d^2 x q \sqrt{2V(\phi_3)} = 4\pi k \langle \sqrt{2V} \rangle_{S^2} \; , \quad \langle \sqrt{2V} \rangle_{S^2} \equiv \frac{\int d\Omega \sqrt{2V}}{\int d\Omega}.
\end{equation}
The static energy of the gauged BPS submodel is ($Q\equiv (1/2) \tilde K_0 = q+\epsilon_{ij}A_i \partial_j \phi_3$)
\begin{equation}
E = \frac{1}{2} E_0 \int d^2 x \left[ \lambda^2 Q^2 +2 \mu^2 V(\phi_3) + \frac{1}{g^2} B^2 \right]
\end{equation}
where $B$ is the magnetic field, and we used the known fact \cite{GPS} that for static finite energy solutions of the gauged baby Skyrme model the electric field must be zero. To find the BPS bound, we now start with a non-negative expression and try to re-express it as the energy minus a topological term, such that the topological term provides the bound. Concretely, we start from
\begin{equation} \label{non-neg}
0 \le \frac{1}{2}E_0 \int d^2 x \left[ \lambda^2 (Q- W'(\phi_3))^2 + \frac{1}{g^2} (B+g^2 \lambda^2 W(\phi_3))^2 \right] .
\end{equation} 
Expanding the squares, we get for the mixed terms $-2\lambda^2 (q + \epsilon_{ij}A_i \partial_j \phi_3) W' +2\lambda^2 W \epsilon_{ij}\partial_iA_j = -2\lambda^2 q W' + 2\lambda^2 \epsilon_{ij}\partial_j (A_i W)$ and, therefore, a total derivative (which integrates to zero) plus a topological term. The remaining contributions from the squares produce the energy provided that the function $W(\phi_3)$ (which has not yet be determined) obeys the first order ODE
\begin{equation} \label{WV-eq}
\lambda^2 W'^2 + g^2  \lambda^4 W^2 =2\mu^2 V.
\end{equation}
Assuming this we find the BPS bound
$
E \ge E_0 \lambda^2 \int d^2 x qW' = 4\pi  E_0 \lambda^2 k \langle W' \rangle_{S^2} ,
$
with equality if the two (first order) BPS equations $Q-W' =0$ and $B+g^2 \lambda^2 W=0$ hold.

\subsection{Remarks}
\begin{itemize}
\item Equation (\ref{WV-eq}) is well-known in supergravity where it provides the relation between potential and superpotential, so we call $W$ the "superpotential" and eq. (\ref{WV-eq}) the "superpotential equation".
\item Eq. (\ref{WV-eq}) is of first order and seems to have a one-parameter family of solutions. This is, however, not the case. At the vacuum value $\vec \phi = \vec \phi_0$, i.e., $\phi_3 =1$ (where $V(1)=0$), $W(\phi_3)$ must take the unique value $W(1)=0$. The solution is, therefore, unique near the vacuum. The global existence of this solution in the whole interval $1\ge \phi_3 \ge -1$ is a different and rather nontrivial problem. 
\item It can be proved that the BPS equations imply the static second order field equations. The existence of nontrivial BPS solitons solutions can therefore be expected.
\item The global existence of the superpotential is a necessary condition for the existence of BPS soliton solutions. For some potentials like, e.g., $V=1-\phi_3$ or $V=(1-\phi_3)^2$ (and probably for the whole family $V=(1-\phi_3)^a$, $a\ge 1$) the superpotential exists globally, whereas for others, like the two-vacua potential $V=1-\phi_3^2$, it doesn't.
\item In all cases where soliton solutions have been calculated numerically, their energies saturate the BPS bound, and these solitons are, therefore, BPS solutions. For potentials without a globally existing superpotential, numerical soliton solutions have not been found. This motivates the conjecture that all solitons are, in fact, BPS solitons.
\item For some potentials, exact superpotentials and exact BPS solitons can be calculated.
\item The gauged BPS baby Skyrme model shares all the infinitely many symmetries and conservation laws with its ungauged version.
\item For more details, see \cite{gaugedBPSbaby}.

\end{itemize}

\subsection{Open problems and further research}
\begin{itemize}
\item For wich potentials the superpotential exists globally? Which further conditions the potential must satisfy such that BPS solitons exist?
\item Can the present construction be generalized to the gauged BPS Skyrme model in 3+1 dimensions?
\item Can the gauged (BPS) baby Skyrme model be supersymmetrized and can the BPS bound we found be related to supersymmetry (we remark that the baby Skyrme model can be supersymmetrized \cite{susy-bS})?
\item Can the novel BPS bound discussed in these notes be used for other field theories? Etc.  

\end{itemize}

\ack
The authors acknowledge financial support from the Ministry of Education, Culture and Sports, Spain (grant FPA2008-01177), 
the Xunta de Galicia (grant INCITE09.296.035PR and
Conselleria de Educacion), the
Spanish Consolider-Ingenio 2010 Programme CPAN (CSD2007-00042), and FEDER. 
CN thanks the Spanish
Ministery of
Education, Culture and Sports for financial support (grant FPU AP2010-5772).
Further, AW was supported by polish NCN grant 2011/01/B/ST2/00464.

\end{document}